\documentclass[pre,twocolumn,showpacs,amssymb]{revtex4}
\usepackage{graphicx}

\begin{document}

\title{Fluctuating Fronts as Correlated Extreme Value Problems: An
Example of Gaussian Statistics}

\author{Debabrata Panja} \affiliation{Institute for Theoretical
Physics, Universiteit van Amsterdam, Valckenierstraat 65, 1018 XE
Amsterdam, The Netherlands}

\date{\today}
\begin{abstract} 
In this paper, we view fluctuating fronts made of particles on a
one-dimensional lattice as an extreme value problem. The idea is to
denote the configuration for a single front realization at time $t$ by
the set of co-ordinates
$\{k_i(t)\}\equiv[k_1(t),k_2(t),\ldots,k_{N(t)}(t)]$ of the
constituent particles, where $N(t)$ is the total number of particles
in that realization at time $t$. When $\{k_i(t)\}$ are arranged in the
ascending order of magnitudes, the instantaneous front position can be
denoted by the location of the rightmost particle, i.e., by the
extremal value
$k_f(t)=\text{max}[k_1(t),k_2(t),\ldots,k_{N(t)}(t)]$. Due to
interparticle interactions, $\{k_i(t)\}$ at two different times for a
single front realization are naturally not independent of each other,
and thus the probability distribution $P_{k_f}(t)$ [based on an
ensemble of such front realizations] describes extreme value
statistics for a set of correlated random variables. In view of the
fact that exact results for correlated extreme value statistics are
rather rare, here we show that for a fermionic front model in a
reaction-diffusion system, $P_{k_f}(t)$ is Gaussian. In a bosonic
front model however, we observe small deviations from the Gaussian.
\end{abstract}


\maketitle

\section{Introduction\label{sec1}}

Extreme value statistics of random variables plays a diverse role in
physics, chemistry and biology \cite{gumbel,galambos,berman}. The
topic concerns the probability distributions of the extrema (i.e., the
maximum $k_{max}$ or the minimum $k_{min}$) of a set of $N$ random
variables $\{k_1,k_2,\ldots,k_N\}$ in the limit
$N\rightarrow\infty$. When the random variables $k_i$ are
uncorrelated, the probability distribution of $k_{min}$ belongs to one
of the three universality classes \cite{mezard}, but the
identification of similar universality classes for the extreme value
statistics of correlated random variables is still largely an open
problem. A few results relating to extreme value statistics for
correlated random variables in physics, computer science and
mathematics have been obtained in the recent past \cite{carp,satya};
nevertheless, any exact result that can be obtained for correlated
random variables is an important addition to the present state of
knowledge.

From this perspective, in this paper, we present two main results
relating to fluctuating fronts made of discrete particles on a
one-dimensional lattice. Before we proceed further with our
formulation of the problem, we must note that an intriguing connection
between the extreme value statistics of correlated random variables
and travelling fronts have already emerged from the recent works
\cite{satya,satya1}. To be more precise, these works have
demonstrated, for the models they studied, that the cumulative
probability distributions of extreme values for correlated random
variables admit propagating front solutions, wherein the variance of
the extremal variable is the front width itself. Our formulation here,
however, is completely the other way round: namely that our systems
consist of many {\it interacting\/} particles, where the dynamics of
the systems {\it already\/} admits front solutions propagating into
unstable states. Although in a deterministic mean-field description,
these fronts propagate with a fixed speed and a fixed shape at long
times, due to the presence of stochasticity involving many particles,
the front in a given realization of the system does not move with a
uniform speed even at long times --- instead, the front speed averaged
over an ensemble of front realizations approaches a constant in time
at long times. Moreover, as a result of the inherent stochasticity in
these systems, the individual front realizations that are initially
aligned with each other do not remain so at a later time; instead
their displacement w.r.t. each other keeps increasing with time (see
Fig. 4 of Ref. \cite{review} for an illustration). As explained below,
it is the dynamics of the individual front realizations in the
ensemble that we pose as a correlated extreme value problem in this
paper.

The correspondence between the extremal value statistics and the
fluctuating fronts in these systems is easily made by first noticing
that in any realization of these systems, the front position can be
denoted by the instantaneous position $k_f(t)$ of the foremost (or the
rightmost) particle  \cite{panja1,panja2}. The interest then lies in
the probability distribution $P_{k_f}(t)$, which describes the
statistics of the front position in time for an ensemble of front
realizations. Secondly, in a snapshot of one single realization, the
configuration of the system is described by the locations of the
particles (as random variables)
$\{k_i(t)\}\equiv[k_1(t),k_2(t),\ldots,k_{N(t)}(t)]$, where $N(t)$ is
the total number of particles in that realization at time $t$. Then
the instantaneous front position $k_f(t)$ in this formulation is then
simply the extremal value
$\text{max}[k_1(t),k_2(t),\ldots,k_{N(t)}(t)]$. Due to the
interparticle interaction within the system defined by the microscopic
rules of the dynamics, $\{k_i(t)\}$ are naturally not independent of
each other, and thus $P_{k_f}(t)$ simply describes the statistics of
the extreme for a set of correlated random variables.

In this paper, we consider two different systems that admit front
solutions propagating into unstable states: (a) a fermionic
reaction-diffusion system A$\leftrightharpoons$A$+$A
\cite{panja2,kerstein1,ba2} in Sec. \ref{sec2}, where we show that
$P_{k_f}(t)$ is Gaussian, and (b) the so-called (bosonic) clock model
\cite{ramses} in Sec. \ref{sec3}, where $P_{k_f}(t)$ has small
deviations from the Gaussian. It is important to note here that the
front solutions in these models have been analyzed before, in the
sense that both the front speed $v=\lim_{t\rightarrow\infty}d\langle
k_f(t)\rangle/dt$ and the front diffusion coefficient
$D_f=\lim_{t\rightarrow\infty}d\langle[k_f(t)-vt]^2\rangle/dt$,
respectively based on the first and the second moments of
$P_{k_f}(t)$, have previously been analyzed and numerically measured
\cite{panja2,kerstein1,ba2,ramses,bd,review}. The higher (than second)
moment of $P_{k_f}(t)$, or $P_{k_f}(t)$ itself, however, have not been
extensively studied before.

The paper is finally ended with a discussion in Sec. \ref{sec4}.

\section{A Fermionic Reaction-Diffusion Model and Gaussian Behaviour
of $P_{k_f}(t)$\label{sec2}}

\begin{figure}[h]
\begin{center}
\includegraphics[width=0.45\textwidth]{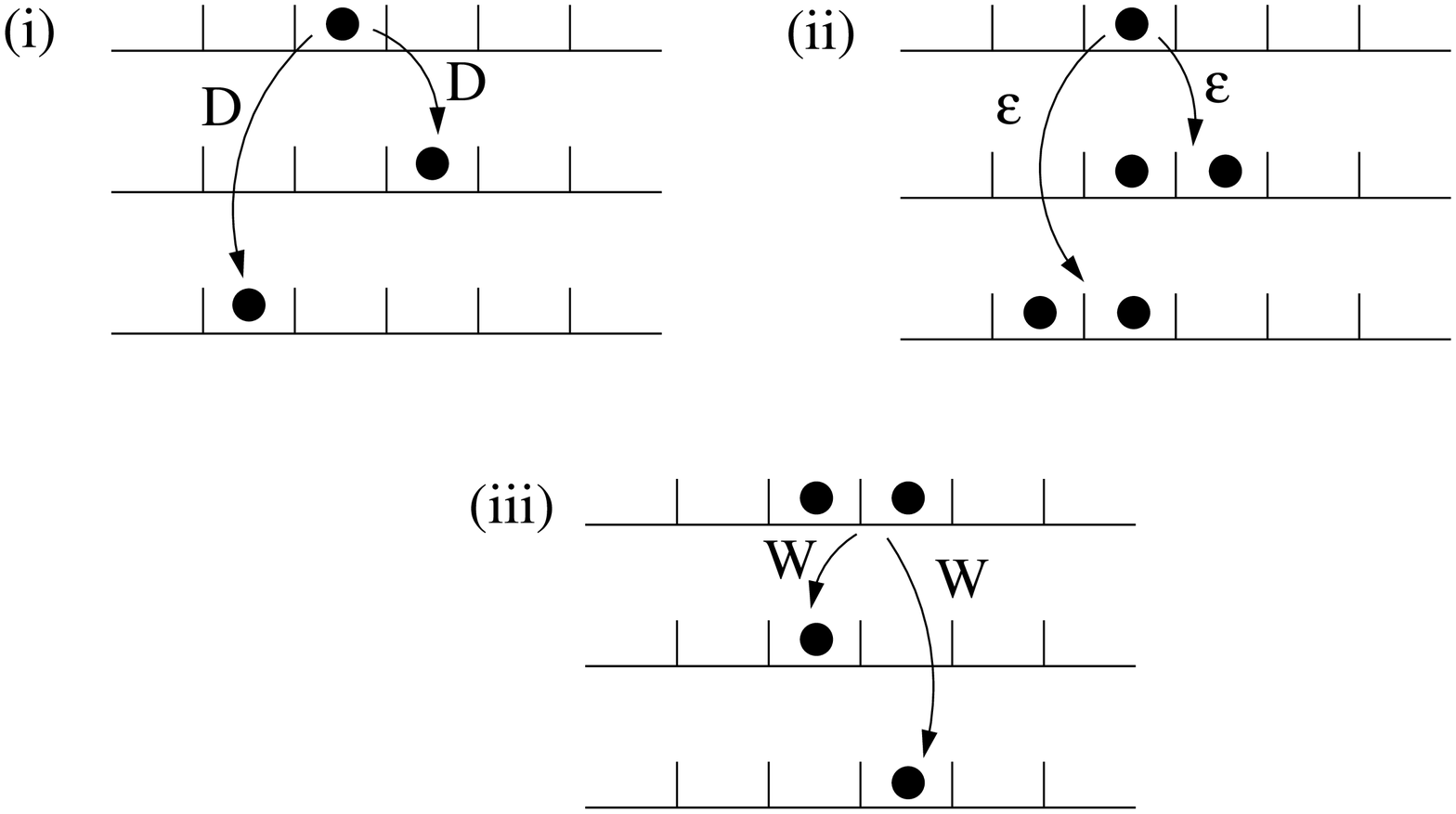}
\end{center}
\caption{The microscopic processes that take place inside the system:
{\em (i)} A  diffusive  hop with rate $D$ to a neighboring empty site;
{\em (ii) } Creation  of a new particle on a site neighboring an
occupied site with rate $\varepsilon$; {\em (iii)} Annihilation  of a
particle on a site adjacent to an occupied site at a rate
$W$. \label{fig1}}
\end{figure}

In this model, we consider a one-dimensional lattice on which at most
one A particle is allowed per lattice site at any instant --- hence
the model is named fermionic. The particles can undergo the following
three basic moves, shown in Fig. \ref{fig1}: (i) A particle can
diffuse to any one of its neighbour lattice sites with a diffusion
rate $D$, provided this neighbouring site is empty. (ii) Any particle
can give birth to another one on any one of its empty neighbour
lattice site with a birth rate $\varepsilon$. (iii) Any one of two A
particles belonging to two neighbouring filled lattice sites can get
annihilated with a death rate $W$.

The lattice indexed by $k$ that we consider in this problem is
semi-infinite. The left boundary is impenetrable --- no particle can
diffuse across the left boundary located on the left of the lattice
site $k=0$, while the system is of infinite extent on the right
side. Following the usual convention, we start with a step initial
condition, i.e., at time $t=0$, there exists a finite
$k_{\text{right}}$, such that all lattice sites $0\le k\le
k_{\text{right}}$ are occupied and $k>k_{\text{right}}$ are
empty. This system then admits a fluctuating (and propagating) front
solution for $t>0$.

Earlier work on models of this type has appeared in
Refs. \cite{kerstein1,ba2,panja2,ba1}. In the general case there are
essentially only two nontrivial parameters in our model, e.g., the
ratios $D/\varepsilon$ and $D/W$, since an overall multiplicative
factor simply sets the time scale. When these ratios tend to infinity,
the front speed approaches its mean field value \cite{kerstein1}.

For an ensemble of front realizations, let us denote the probability
distribution for the foremost occupied lattice site to be at lattice
site ${k_f}$ at time $t$ by $P_{k_f}(t)$. The evolution of
$P_{k_f}(t)$ is then described by
\begin{eqnarray}
\frac{dP_{k_f}}{dt}\,=\,(D\,+\,\varepsilon)\,P_{{k_f}-1}\,+\,\left[D\,P^{\mbox{\scriptsize
empty}}_{{k_f}+1}\,+\,W\,P^{\mbox{\scriptsize
occ}}_{{k_f}+1}\right]\nonumber\\&&\hspace{-6.9cm}-\,(D\,+\,\varepsilon)\,P_{{k_f}}\,-\,\left[D\,P^{\mbox{\scriptsize
empty}}_{{k_f}}\,+\,W\,P^{\mbox{\scriptsize occ}}_{{k_f}}\right]\,.
\label{e2}
\end{eqnarray}
Here $P^{\mbox{\scriptsize occ}}_{{k_f}}(t)$ and $P^{\mbox{\scriptsize
empty}}_{{k_f}}(t)$ respectively denote the joint probabilities that
the foremost particle is at site $k_f$ and that the site  $k_f-1$ is
occupied and empty. Clearly, $P_{k_f}(t)=P^{\mbox{\scriptsize
occ}}_{{k_f}}(t)+P^{\mbox{\scriptsize empty}}_{{k_f}}(t)$, and
$\sum_{k_f}P_{k_f}(t)=1$. The first term on the r.h.s. of
Eq. (\ref{e2}) describes the increase in $P_{k_f}(t)$ due to the
advance of a foremost occupied lattice site from position
${k_f}$$-$$1$, while the second term describes the increase in
$P_{k_f}(t)$ due to the retreat of a foremost occupied lattice site
from position ${k_f}$$+$$1$. The  third and the fourth terms
respectively describe the decrease in $P_{k_f}(t)$ due to the advance
and retreat of a foremost occupied lattice site from position
${k_f}$. It is clear from this formulation that the dynamics of
$P_{k_f}(t)$ is effectively obtained only from the coupled interaction
between the foremost particle and the site just behind it.

In addition to Eq. (\ref{e2}), we have
\begin{equation}
P^{\mbox{\scriptsize occ}}_{k_f}(t)=\rho_{k_f-1}(t) P_{k_f}(t),
\label{neweq}
\end{equation}
where $\rho_{k_f-1}(t)$ is the conditional probability of having the
$({k_f}$$-$$1)$th lattice site occupied. At large $t$,
$\rho_{k_f-1}(t)$ should be independent of $k_f$ and $t$, and one can
replace $\rho_{k_f-1}(t)$ by $\bar\rho$ in Eq. (\ref{neweq}), where
the numerical value of $\bar\rho$ depends only on those of $D$,
$\varepsilon$ and $W$. Similarly, the set of (time and
$k_f$-independent) conditional occupation densities $\rho_{k_f-m}(t)$
for $m\ge1$ can be thought of as determining the front profile in a
frame moving with the foremost particle of each front realization (see
Fig. 5 of Ref. \cite{panja2} for an illustration).

With the condition $P_{k_f}(t)=P^{\mbox{\scriptsize
occ}}_{{k_f}}(t)+P^{\mbox{\scriptsize empty}}_{{k_f}}(t)$, and the
notation $q=\bar\rho(W-D)$, at large $t$, Eq. (\ref{e2}) can be
rewritten as
\begin{eqnarray}
\frac{dP_{k_f}}{dt}=\frac{1}{2}\,(2D+\varepsilon+q)\,\left[P_{{k_f}+1}+P_{{k_f}-1}-2P_{k_f}\right]\nonumber\\&&\hspace{-4.9cm}-\,\frac{1}{2}\,(\varepsilon-q)\,\left[P_{{k_f+1}}-P_{{k_f-1}}\right]\,,
\label{e3}
\end{eqnarray}
which is clearly a diffusion equation for $P_{k_f}(t)$ with a
drift. After having aligned the locations of the foremost particles
for all realizations in the ensemble, say at $k_f=k_{\text{in}}$ at
time $t_{\text{in}}\gg1$ [i.e.,
$P_{k_f}(t_{\text{in}})=\delta_{k_f,k_{\text{in}}}$], we are
interested in the solution of $P_{k_f}(t)$. In fact, Eq. (\ref{e3})
can be solved by taking a discrete Fourier transform in $k_f$, but due
to the redundancy of the wavevector modulo any multiple of $2\pi$, the
magnitude of the wavevector has to be kept confined only within the
first Brillouin zone $(-\pi,\pi]$. Then for $\Delta
t=t-t_{\text{in}}\gg1$, it is easily seen that the dominant
contribution to $P_{k_f}(t)$ comes from the wavevector in the
neighbourhood of zero, yielding \cite{same}
\begin{eqnarray}
P_{k_f}(t)=\frac{\exp\left[-\displaystyle{\frac{(k_f-k_{\text{in}}-v\Delta
t)^2}{4D_f\Delta t}}\right]}{\sqrt{4\pi D_f\Delta t}}\,.
\label{e4}
\end{eqnarray}
Here, $v=\varepsilon-q$ is the front speed and $D_f=2D+\varepsilon+q$
is the front diffusion coefficient, as already derived as the first
and the second moment of $P_{k_f}(t)$ in Ref. \cite{panja2}.

\section{The Clock Model and the Non-Gaussian Behaviour of
$P_{k_f}(t)$\label{sec3}}

The clock model was originally invented in the context of the largest
Lyapunov exponent for a gas of hard spheres \cite{ramses}. In this
model, one considers a system of $N$ clocks with integer readings
$\{k_i\}$. The dynamics of the clocks involve  binary ``collisions''
between any two randomly chosen clocks in continuous time. In a
collision between two clocks with pre-collisional readings $k_i$ and
$k_j$, the post-collisional readings of both clocks are updated to
$\text{max}[k_i,k_j]+1$.

In the clock reading space, which can be imagined as a one-dimensional
lattice, the number of clocks $N_k$ with readings $k$ or higher for
any realization of the clock model admits a fluctuating (and
propagating) front solution \cite{ramses}. Clock model allows more
than one clock with the same reading and hence the model is
bosonic. Conventionally, all clock readings in any realization are
initially (i.e., at $t=0$), taken to be equal to zero --- for the
propagating front, this corresponds to the step initial condition
\cite{ramses}. Once again we denote the largest clock reading in any
realization at time $t$ by $k_f(t)$.

In the deterministic mean-field limit, the propagating front in the
clock model is a pulled front \cite{ute}, and if the time is rescaled
in order to have the mean collision frequency of a single clock equal
to unity, the front propagates with a speed $v^*=4.31107\ldots$
\cite{ramses}. However, due to stochasticity effects associated with
discreteness effects of the clocks and their readings, in the limit of
asymptotically large $N$, the front speed $v$ and front diffusion
coefficient $D_f$, which could be measured following the procedure
described in the last paragraph of Sec. \ref{sec1}, have the property
that $(v^*-v)\!\propto\!1/\ln^2N$ and $D_f\propto1/\ln^3N$
\cite{bd}. Thus, the clock model is an example of a fluctuating
``pulled'' front \cite{review,panja1,bd}.

To write a master equation for $P_{k_f}(t)$ defined over an ensemble
in the clock model, it may be argued that the reading of any clock in
any realization can only increase with time; and thus, $P_{k_f}(t)$
can increase when in a realization, one of the clocks with largest
reading $k_f-1$ is involved in a collision with another
one. Similarly, $P_{k_f}(t)$ can decrease when in a realization, one
of the clocks with largest reading $k_f$ is involved in a collision
with another one. If we now denote the conditional probability of the
number of clocks with largest reading $k_f$ to be $n_{k_f}(t)$ at time
$t$ by ${\cal P}(n_{k_f},t)$, the master equation for $P_{k_f}(t)$
reads
\begin{eqnarray}
\frac{dP_{k_f}}{dt}=\left[\sum_{n_{k_f-1}}C(n_{k_f-1})\,{\cal
P}(n_{k_f-1},t)\right]P_{k_f-1}\nonumber\\&&\hspace{-4.5cm}-\,\left[\sum_{n_{k_f}}C(n_{k_f})\,{\cal
P}(n_{k_f},t)\right]P_{k_f}\,.
\label{e5}
\end{eqnarray}
Here, $C(n_{k_f})$ is the rate of collisions that involve a clock with
reading $k_f$ for a realization with $k_f$ as the largest of the clock
readings. From Eq. (\ref{e5}), one might now further argue that at
large $t$, the quantities within the large square brackets in
Eq. (\ref{e5}) are independent of $t$ and $k_f$, and thus at large
$t$, Eq. (\ref{e5}) should reduce to a form
$\displaystyle{\frac{dP_{k_f}}{dt}=\bar c\,[P_{k_f-1}\,-\,P_{k_f}]}$,
where $\bar c=\displaystyle{\left[\sum_{n_{k_f}}C(n_{k_f})\,{\cal
P}(n_{k_f},t)\right]}$ at large $t$. However, for any finite value of
$N$, the simple-minded replacement of
$\displaystyle{\left[\sum_{n_{k_f}}C(n_{k_f}){\cal
P}(n_{k_f},t)\right]}$ by a $t$ and $k_f$-independent quantity $\bar
c$ in Eq. (\ref{e5}) at large $t$ is incorrect for the clock model ---
caused by the fact that ${\cal P}(n_{k_f},t)$ does {\it not\/} become
time-independent at large $t$ \cite{cannot} --- as we argue below.

The observation we make, in order to argue that ${\cal P}(n_{k_f},t)$
does not become independent of $t$ at large $t$, is that the largest
clock reading in any realization does not increase smoothly in time
with a rate $v$ even at large $t$. Instead, after attaining a new
integer value, the largest of the clock readings for any given
realization does not change for some time-interval (hereafter denoted
by $\delta t$) of typical magnitude $1/v$ before attaining the next
integer value \cite{similar}. Generally speaking, during any of these
time-intervals, the number of clocks with the largest clock reading in
any realization increases with time; and the number of clocks with the
largest reading at any instant in a given realization depends on how
long the largest clock reading remains unchanged at its value. The
conditional probability ${\cal P}(n_{k_f},t)$ can thus be written as
\begin{eqnarray}
{\cal P}(n_{k_f},t)\,=\,\int_{0}^{\infty}d(\delta t)\,\wp_1(\delta
t,t)\,\wp_2(n_{k_f},\delta t,t)\,,
\label{enew1}
\end{eqnarray}

With Eq. (\ref{enew1}) in the back of our minds, we now return to the
statement to the second sentence of Ref. \cite{similar}: namely that
front propagation in any realization of the clock model is coded in
the {\it sequential\/} values of the time-intervals $\{\delta t_i\}$
between the consecutive changes of the largest clock reading. In this
description, the point to note is that the $\delta t_i$ values are
very strongly correlated with each other; e.g., a large $\delta t$ is
almost always followed by several small values of $\delta t$ and vice
versa (the large or smallness of $\delta t$ are decided in comparison
to $1/v$) \cite{review,panja1}. Due to such strong dependence of of
the $\delta t$ values on the evolution histories of individual
realizations, it is easily conceivable that the shape of the
probability distribution $\wp_1(\delta t,t)$ lacks $t$-independence at
large $t$.
\begin{figure}[!h]
\begin{center}
\includegraphics[width=0.35\textwidth,angle=270]{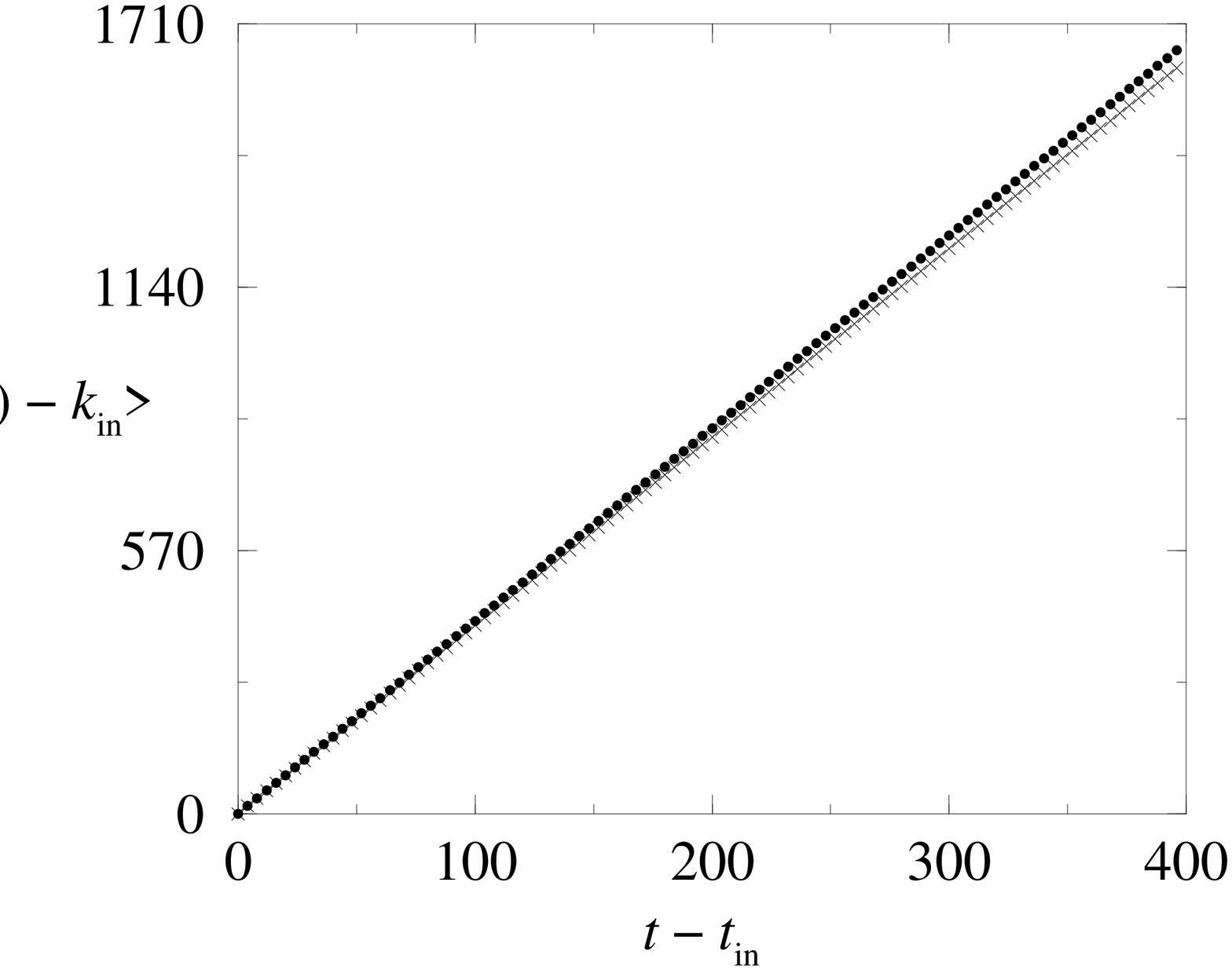}
\end{center}
\begin{center}
\includegraphics[width=0.33\textwidth,angle=270]{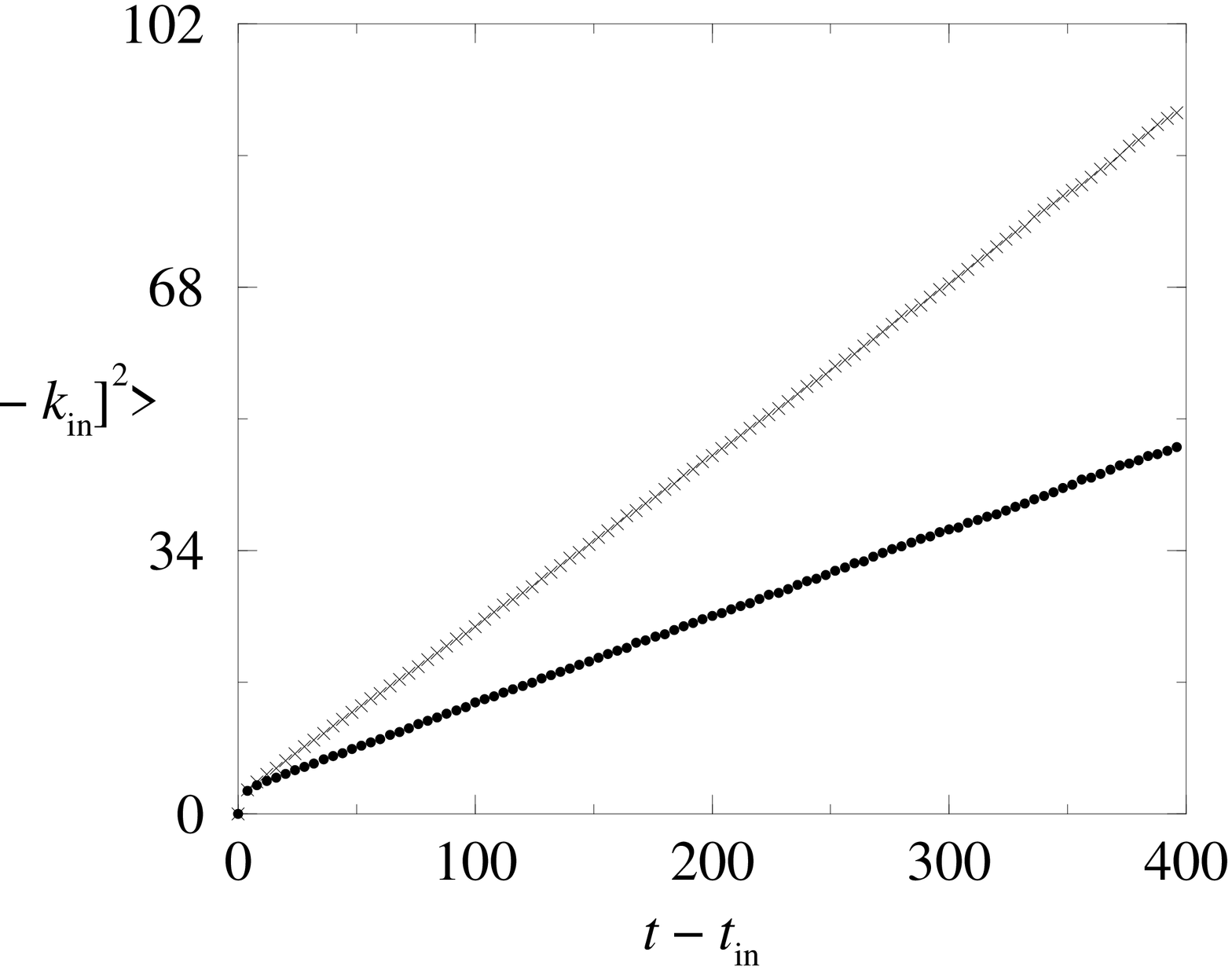}
\end{center}
\caption{Top figure: Simulation results for $\langle
k_f(t)-k_{\text{in}}\rangle$ as a function of
$t-t_{\text{in}}$. Bottom figure: Simulation results for
$\langle[k_f(t)-k_{\text{in}}]^2\rangle$ as a function of
$t-t_{\text{in}}$. Apart from an initial transient for
$t-t_{\text{in}}\lesssim 10$, $\langle[k_f(t)-k_{\text{in}}]^2\rangle$
increases linearly with $t$, indicating that the front wandering is
diffusive. Crosses correspond to $N=10^4$ (front speed $v=4.08$ and
front diffusion coefficient $D_f=0.112$) and filled circles correspond
to $N=10^5$ (front speed $v=4.17$ and front diffusion coefficient
$D_f=0.056$) in both figures. \label{fig2}}
\end{figure}
where $\wp_1(\delta t,t)$ is the probability that the largest clock
reading became $k_f$ at time $(t-\delta t)$ and remains so until time
$t$, and $\wp_2(n_{k_f},\delta t)$ is the probability of having
$n_{k_f}$ clocks at time $t$ when the largest clock reading became
$k_f$ at time $(t-\delta t)$ and remains so until time $t$. Using
Eq. (\ref{enew1}), the $t$-dependence of ${\cal P}(n_{k_f},t)$ can
then be argued {\it in terms of the $t$-dependences of $\wp_1(\delta
t,t)$ and $\wp_2(n_{k_f},\delta t,t)$}.

The $t$-dependence of $\wp_2(n_{k_f},\delta t,t)$ can be argued in a
similar way. In realizations for which the largest clock value became
$k_f$ at time $(t-\delta t)$ and remains so until time $t$, how many
clocks share the largest clock reading at time $t$ depends on the
time-dependence of the number of clocks $n_{k_f-1}$ with clock
readings $(k_f-1)$ between times $(t-\delta t)$ and $t$ --- after all,
any clock that attains a reading $k_f$ must come out of a collision
that involves a clock with reading $(k_f-1)$. Between times $(t-\delta
t)$ and $t$, $n_{k_f}-1$ changes also with time, and thus the
probability distribution $\wp_2(n_{k_f},\delta t,t)$ inherently
connects to the {\it fluctuations in the  shapes of individual front
realizations\/} \cite{review,bd}. These fluctuations have a typical
correlation time $\propto\ln^2N$ \cite{review,bd,panja3}. For
$N\rightarrow\infty$, this correlation time also becomes large, and
one therefore expects the shape of $\wp_2(n_{k_f},\delta t,t)$ to also
depend on $t$ via the strong dependence of $n_{k_f-1}$ on the
evolution histories of individual realizations at earlier times.
\begin{figure}[!h]
\begin{center}
\includegraphics[width=0.4\textwidth,angle=270]{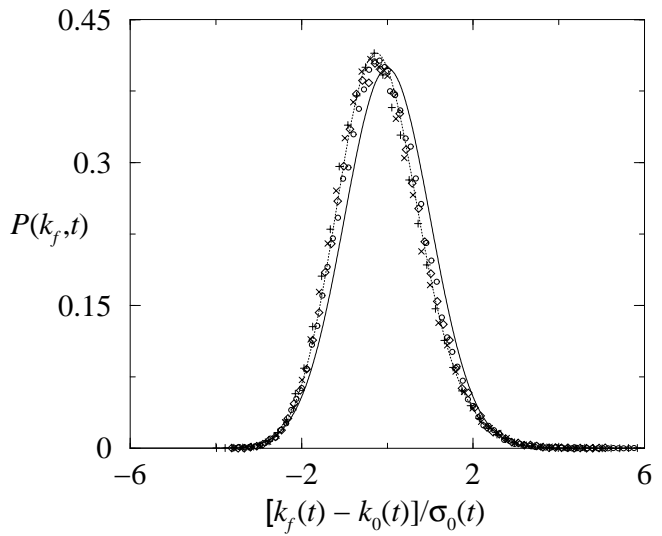}
\end{center}
\caption{$P_{k_f}(t)$ for the clock model; pluses: data for $N=10^4$
at $t=t_{\text{in}}+99$, circles: data for $N=10^4$ at
$t=t_{\text{in}}+297$, crosses: data for $N=10^5$ at
$t=t_{\text{in}}+198$ and diamonds: data for $N=10^5$ at
$t=t_{\text{in}}+396$. Solid line: normalized Gaussian distribution
with mean zero and variance unity. Dotted line: numerically obtained
curve for the collapsed data.\label{fig3}}
\end{figure}

With no further simplification of Eq. (\ref{e5}) possible, let alone
an exact solution for $P_{k_f}(t)$ like in Eq. (\ref{e4}), we can only
study $P_{k_f}(t)$ for the clock model only via simulation. Our
(molecular dynamics) simulation methods are as follows: we choose an
ensemble of ${\cal N}=50000$ realizations of $N=10^4$ clocks and set
all clock values zero at $t=0$. We then let each realization evolve
until time $t_{\text{in}}=800$ units. At $t_{\text{in}}$, we align the
different realizations in such a way that the largest of all the clock
values coincide at $k=k_{\text{in}}$. We then follow the locations of
the largest clock values for each realization until
$t-t_{\text{in}}=400$. We also repeat the calculations for the same
values of ${\cal N}$, $t_{\text{in}}$ and $k_{\text{in}}$ but for
$N=10^5$. The ensemble average $\langle k_f(t)-k_{\text{in}}\rangle$
and $\langle[k_f(t)-k_{\text{in}}]^2\rangle$ for ${\cal N}=50000$ as a
function of $(t-t_{\text{in}})$ both for $N=10^4$ and $10^5$ have been
shown in Fig. \ref{fig2}.

To obtain $P_{k_f}(t)$ numerically from the above data, we now proceed
in the following way. First we select two different time instants for
each value of $N$ to take snapshots of the entire ensemble of
$k_f$-values: for $N=10^4$, we choose $t=t_{\text{in}}+99$ and
$t_{\text{in}}+297$, and for $N=10^5$, we choose $t=t_{\text{in}}+198$
and $t_{\text{in}}+396$. Having used the best fit method from the data
of Fig. \ref{fig2}, we then identify the location of the mean front
position $k_0(t)$ and the standard deviation $\sigma_0(t)$
[$\sigma_0(t)$ effectively behaves as
$\sim\sqrt{2D_f(t-t_{\text{in}})}$ as seen in the bottom plot of
Fig. \ref{fig2} for large $(t-t_{\text{in}})$], for two different
values of $N$ at these different time instants. Finally, with the
histograms ${\cal N}(k_f,t)/[{\cal N}\sigma_0(t)]$  plotted as a
function of $[k_f-k_0(t)]/\sigma_0(t)$, where ${\cal N}(k_f,t)$ is the
number of realizations with largest of the clock values $k_f$ at time
$t$, we expect a good data collapse, and the corresponding curve then
gives us the normalized $P_{k_f}(t)$. Notice that the procedure that
we followed to obtain $k_0(t)$ and $\sigma_0(t)$ [and subsequently the
numerical curve for $P_{k_f}(t)$] at the above time instants does not
guarantee $\langle k_f(t)\rangle-k_0(t)\equiv0$ and $\langle
[k_f(t)-k_0(t)]^2\rangle\equiv\sigma_0^2(t)$; instead, the $\langle
k_f(t)\rangle-k_0(t)$ and the $\langle
[k_f(t)-k_0(t)]^2\rangle/\sigma_0^2(t)$ values are in fact very close
to zero and unity respectively.

This data collapse is shown by means of the numerically obtained
dotted curve in Fig. \ref{fig3}. {\it Further analysis of the data
(not presented here) clearly shows that the dotted curve does not
belong to any of the known universality classes for the extreme value
statistics of uncorrelated random variables \cite{mezard}; instead, it
appears to resemble the normalized Gaussian distribution rather
closely}. To facilitate comparison, we therefore plot $P_{k_f}(t)$
against the normalized Gaussian distribution (with mean zero and
variance unity). It is clear from Fig. \ref{fig3} that the dotted
curve is positively skewed; direct measurement of the third cumulant
from the data also confirms this positive skewness behaviour of
$P_{k_f}(t)$. The most noteworthy feature is the longer right tail of
the collapsed data than the left tail, implying that the probability
for large positive deviation around the mean for the clock model is
larger than that of large negative deviation. This is indeed
consistent with positively skewed $P_{k_f}(t)$ --- as stated before,
$\langle k_f(t)-k_0(t)\rangle\simeq0$ for all snapshots.

While Fig. \ref{fig3} certainly provides an example of deviation from
Gaussian statistics when the fluctuating front propagation is seen as
a correlated extreme value problem, it also provides an interesting
perspective from the point of view of fluctuating front propagation
literature. As already mentioned before, clock model is an example of
fluctuating ``pulled'' fronts, and the expression for $v$ and the
scaling for $D_f$ due to the discrete particle stochasticity effects
in the limit of asymptotically large values of $N$ are known for the
last few years. It is also known that over a time interval $\Delta t$
at large $t$, the second moment of $P_{k_f}(t)$, i.e.,
$\langle[k_f(\Delta t)-k_{\text{in}}-v\Delta t]^2\rangle\sim2D_f\Delta
t$ for all values of $N$. Figure \ref{fig3} however shows that the
information regarding the second moment is clearly not enough to
characterize $P_{k_f}(t)$. Nevertheless, the data collapse shows that
at large $t$, $P_{k_f}(t)\equiv P[(k_f-k_{\text{in}}-v\Delta
t)/\sqrt{2D_f\Delta t}]/\sqrt{2D_f\Delta t}$ (i.e., the dotted line in
Fig. \ref{fig3}) is a characteristic curve for the clock model, and
this characteristic curve is not Gaussian for the values of $N$
studied here. The statement that ``the front wandering is diffusive''
at any value of $N$ for the clock model, therefore, has to be
interpreted only in the sense that the second moment of $P_{k_f}(t)$
increases linearly with time at large $t$ for any value of $N$.

Whether the deviation of the dotted line from the Gaussian is due to
the fact that we have not used extremely large values of $N$ is
however not clear. It is well known that to observe the $1/\ln^2N$
scaling of $(v^*-v)$ and the $1/\ln^3N$ scaling of $D_f$ for
fluctuating ``pulled'' fronts one needs to take $N$ extremely high
\cite{ramses,bd}. Direct molecular dynamics simulations of the clock
model for $N\gtrsim10^6$ are prohibitively slow. The existing
simulation methods at much higher values of $N$ are not only quite
intricate, but they also do not follow the exact dynamics of the model
for all clocks. This particular point, therefore, is left here for
further investigation in future.

\section{Discussion\label{sec4}}

In this paper, we have analyzed front propagation in discrete particle
systems on a one-dimensional lattice as extreme value problems. In
these systems, the positions of the particles can be thought of as
random variables, and these random variables under consideration are
obviously strongly correlated with each other. We have seen that in
the case of the fermionic reaction-diffusion model, the extreme value
problem follows Gaussian statistics. It clearly does not belong to any
of the classes pertaining to extreme value statistics of uncorrelated
random variables. For the (bosonic) clock model however, we see that
the extreme value statistics has a small deviation from the Gaussian,
and additional analysis (now presented here) also clearly shows that
the probability distribution does not belong to any of the known
universality classes for the extreme valus statistics of uncorrelated
random variables. However, due to the unavailability of any analytical
tool, the characterization of this distribution has proved
elusive. Whether the small deviation from the Gaussian is caused by
the fact that we have not used extremely high values of $N$ for our
simulations thus remains an open question.

It is a pleasure to thank Satya N. Majumdar (for pointing out the
connection between fluctuating fronts made of discrete particles on a
lattice and extreme value problems in the first place!), Henk van
Beijeren and Ramses van Zon for many useful discussions. Financial
support was provided by the Dutch Research Organization FOM
(Fundamenteel Onderzoek der Materie) and in part by the French research
organization CNRS during a short visit to Universit\'{e} Paul Sabatier
in Toulouse, France.


\begin{thebibliography}{03}
\bibitem{same} For $D=W$, Eq. (\ref{e4}) has already been derived in
Ref. \cite{ba2}.
\bibitem{similar} A similar situation for the reaction-diffusion
system X$\leftrightharpoons2$X on a one-dimensional lattice has been
analyzed in detail in Ref. \cite{panja1}. In fact, front propagation
for a given realization in any of these discrete particle and lattice
models is effectively coded in the (strongly correlated) sequence
$\delta t_1,\delta t_2,\ldots$. Their average magnitude is precisely
$1/v$, the time required for the front to move forward one lattice
site.
\bibitem{cannot} It is easily seen that the usage of $\bar c$ in
Eq. (\ref{e5}) can only generate a front speed $d\langle
k_f\rangle/dt=\bar c$ and a front diffusion coefficient
$\displaystyle{\frac{1}{2}\frac{\langle k^2_f-\langle
k_f\rangle^2\rangle}{dt}}=\bar c/2$. This clearly violates the known
results for the clock model; namely that for large $N$,
$(v^*-v)\propto1/\ln^2N$ and $D_f\sim1/\ln^3N$ .
\end{thebibliography}
\end{document}